\documentclass[]{spie}
\usepackage[utf8]{inputenc}
\usepackage{lineno}

\usepackage{amsmath,amsfonts,amssymb}
\usepackage{graphicx}
\usepackage[colorlinks=true, allcolors=blue]{hyperref}
\usepackage{lineno}
\usepackage{gensymb}
\usepackage{xcolor}
\usepackage[normalem]{ulem}
\usepackage{orcidlink}

\title{An upgraded frequency-selectable laser source (FLS) calibrator for CMB bandpass characterization}

\author[a]{Lauren J. Saunders\,\orcidlink{0000-0001-6367-6380}}
\author[a, b]{Sara M. Simon\,\orcidlink{0000-0001-9221-7802}}
\author[c]{Shreya Sutariya\,\orcidlink{0000-0002-6971-8809}}
\author[b]{Elisa Russier\,\orcidlink{0009-0005-3268-1044}}
\author[b]{Sanah Bhimani\,\orcidlink{0000-0002-9763-1663}}
\author[d]{Erin Healy\,\orcidlink{0000-0002-3757-4898}}
\author[a, b, c, d, e]{Jeffrey McMahon}

\affil[a]{Fermi National Accelerator Laboratory, Batavia, IL 60510, USA}
\affil[b]{Department of Astronomy and Astrophysics, University of Chicago, 5640 S. Ellis Ave., Chicago, IL 60637}
\affil[c]{Department of Physics, University of Chicago, 5720 S. Ellis Ave., Chicago, IL 60637}
\affil[d]{Kavli Institute for Cosmological Physics, University of Chicago, 5640 S. Ellis Ave., Chicago, IL 60637}
\affil[e]{Enrico Fermi Institute, University of Chicago, 933 E. 56th St, Chicago, IL 60637}

\authorinfo{Corresponding author: \href{mailto:lauren1@fnal.gov}{lauren1@fnal.gov}\\Fermilab Report Number: FERMILAB-CONF-26-0442-LDRD-PPD}
\begin{document}

\maketitle

\abstract
One of the biggest challenges for Cosmic Microwave Background (CMB) experiments comes from the uncertainty in instrument bandpass calibration. Uncertainties in bandpass can limit foreground removal and spectral fitting, which are critical for inflationary and galaxy cluster measurements. CMB experiments currently use Fourier Transform Spectrometers (FTSes) to measure instrument bandpasses. However, FTS systems are currently systematics-limited, so significant improvements in bandpass measurements require novel calibrators. To this end, we developed a Frequency-selectable Laser Source (FLS) calibrator, which uses a laser with adjustable frequency coupled to a system that allows for laser power attenuation. Following initial testing with the first FLS prototype, we developed an upgraded version of the calibrator with improved performance. We present the upgrades to the FLS calibrator, the characterization of the upgraded calibrator and new laser source, and plans for testing with microwave instruments in the field.
\endabstract{}

\keywords{Cosmic Microwave Background, calibrator, bandpass, cosmology, CMB, FLS, FTS}

\section{Introduction}\label{sec:intro}

The Cosmic Microwave Background (CMB) is one of the most powerful datasets for studying the physics of the universe, providing opportunities for insights into the biggest questions in cosmology, including the energy scale of inflation and characteristics of the dark-energy driven expansion of the universe \cite{simons_2018}. Modern experiments require high-sensitivity measurements of the mm-wavelength sky at multiple frequencies, which allows for the separation of the CMB signal from foregrounds, including those created by galactic sources. This foreground separation is dependent on our understanding of the instrument bandpass, and uncertainty in the bandpass introduces systematic contributions to the Sunyaev-Zeldovich (SZ) galaxy cluster power spectrum and the tensor-to-scalar ratio $r$ used to characterize inflation\cite{Giardiello_2024}.

Currently, the standard calibrator used in characterizing instrument bandpasses for mm-wavelength instruments is a Fourier-Transform Spectrometer (FTS). FTSes have achieved an accuracy of $\sim$1-3\%\cite{tommy-fts-advact-2024, alford_fts}, but further improvement with this technique requires significant effort and would likely be incremental. To achieve their scientific goals, current and upcoming experiments, including Simons Observatory (SO) and SPT-3G+, require bandpass calibration to the sub-percent level ($\sim$0.1\%)\cite{Giardiello_2024, Abitbol_2021}.

\begin{figure}
    \centering
    \includegraphics[width=\linewidth]{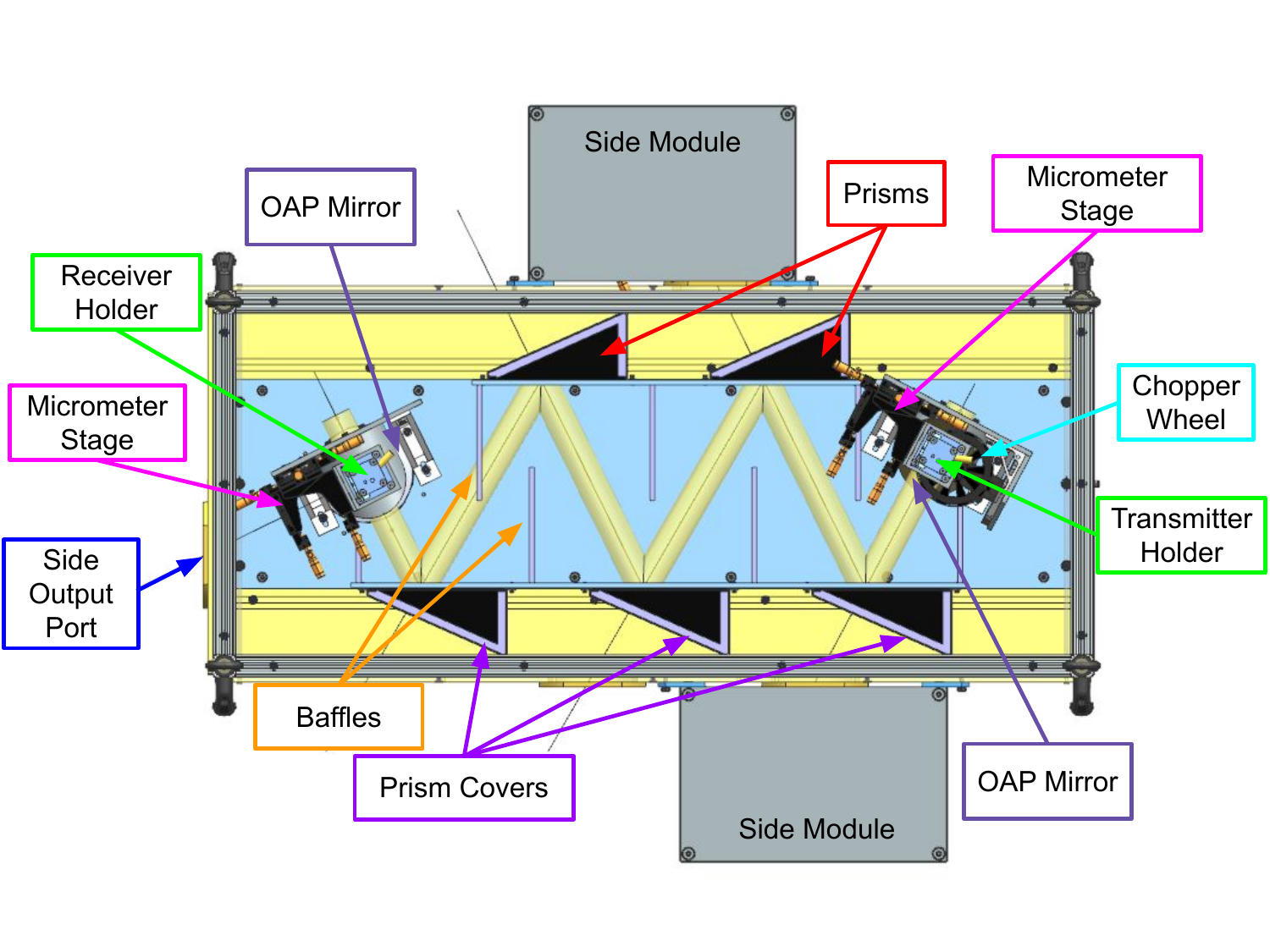}
    \caption{A drawing of the Version 2 FLS in the self-calibration setup as seen from the top. The transmitter photomixer is mounted on the right-side micrometer stage, facing down toward an off-axis parabolic (OAP) mirror, which reflects the light at a 90$^\circ$ angle and collimates the beam. The light from the laser is reflected off of 5 prisms at a 30$^\circ$ angle. A second OAP mirror is positioned after the end of the series of prisms, reflecting the beam upward and focusing it onto the receiver photomixer, which is mounted on the left-side micrometer stage. Between the lasers, narrow baffles lined with HR-10 are installed to prevent errant reflections; the first and last baffles in the series have partial circles cut out to accommodate the geometry of the calibrator. Each of the prisms is surrounded by an aluminum enclosure lined with HR-10. Also shown are two side modules, which can be added behind the prisms in three positions to sample some of the laser signal that passes through the prisms. These side modules can be used to house a power meter and/or spectrometer to characterize the input signal in real time.}
    \label{fig:fls-from-top}
\end{figure}

To achieve this level of accuracy in our bandpass measurements, we are developing the Frequency-selectable Laser Source (FLS) calibration system\cite{shreya-so-fls-2022-poster, shreya-so-fls-2024-poster, shreya-fls-optics-2026}, which, when used in concert with the FTS, is expected to reduce uncertainties to the $\sim$0.1\% level. The FLS (Figure \ref{fig:fls-from-top}) utilizes an off-the-shelf TeraScan 1550 terahertz laser system with coverage from $\sim$20-880 GHz and $<$10 MHz resolution made by Toptica Photonics SE\cite{TeraScan1550}. The source is mounted on mircometer stages in an assembly containing prisms to neutrally attenuate the laser signal and off-axis parabolic (OAP) mirrors to collimate and direct the beam through the series of prisms. Prisms in the system can be substituted for mirror flats to tune the level of attenuation. The assembly also contains baffling and absorptive foam (Eccosorb HR-10) to reduce stray light. In its self-calibration mode shown in Figure \ref{fig:fls-from-top}, the FLS output is directed to a receiver with an OAP mirror. In its measurement mode, the output is instead directed to coupling optics, and the transmitter is chopped with a chopper. The removable prisms and source resolution enable the FLS to measure the edges of the bands with much higher signal to noise and higher resolution when compared with the FTS. Additionally, when measuring the response outside of the intended frequency band, prisms can be removed to increase the power so that smaller responses are visible; through this mechanism, the FLS is particularly useful for searching for out-of-band signal like blue leaks. The FLS and FTS use independent methods of performing bandpass measurements; when used in combination, they are complementary calibrators, allowing us to constrain and mitigate systematics in both systems. The fine frequency resolution can also capture effects that could be missed by sparser FTSes, which is particularly important for future microwave spectrometer experiments that use narrow-band ($\sim$1 GHz) detectors.

The first prototype FLS (Version 1)\cite{shreya-fls-optics-2026} was as a technological pathfinder for this novel calibrator. The Version 1 prototype was tested in the laboratory setting. After the instrument itself was initally characterized, it was used for passband measurements in the Large Aperture Telescope (LAT) optics tube test setup\cite{sierra2025}. These tests demonstrated that the FLS is able to make high precision measurements of the instrument bandpasses. Building on the lessons learned from the initial testing of Version 1, we developed an improved design, Version 2, for use in the next phase of FLS development and testing.

In these proceedings, we detail the upgrades made to the FLS instrument in the Version 2 design (Section \ref{sec:upgrades}) and the in-lab characterization of the Version 2 system (Section \ref{sec:characterization}). We additionally show the optical coupling and software designs used for the first FLS field test with the SO LAT, and show initial measurements of the SO LAT detectors' spectral response (Section \ref{sec:so-lat}).

\section{Instrument Upgrades}\label{sec:upgrades}
The Version 2 prototype was designed to improve on the Version 1 design in both performance and versatility. It features built-in baffling and prism covers to reduce stray light, improved positional tolerances of the optical components, improved capabilities for multiple setups, and improved portability. We have also identified and implemented procedural changes that have improved performance, including a warm-up period and integrated data acquisition.

\subsection{Mechanical Improvements}\label{subsec:mechanical-improvements}

\paragraph{Built-in baffling and prism covers} FLS Version 1 did not initially have baffles or prism covers to block stray reflections. Because they were added after the initial design was finalized, they were not integrated into the design and thus did not have the full desired tolerances and effectiveness. FLS Version 2 is designed to integrate both baffles and prism covers into the design, enabling improved alignment of the components and of the system as a whole.

\paragraph{Positional tolerances} We performed detailed ray trace studies using the \texttt{pyoptools}\cite{pyoptools} package to determine the required positional tolerances within the system. These studies showed that the maximum allowed tilt of the prisms was $0.018^\circ$, with a maximum allowed tilt of the OAP mirror of $0.9^\circ$. To reach these stringent tolerances, we redesigned the full main assembly. The baffles are used to minimize the tilt in the sidewalls where the prisms are mounted, and countersunk screws enable improved alignment in the main assembly components. The OAP mirrors are aligned using a pin hole. We also developed improved mounting of the transmitter and receiver photomixers. In Version 1, XYZ micrometer stages that screwed directly into the source were used, but the stages could not lock. Both the direct screwing mechanism and the lack of locking resulted in drifts in the alignment on the same timescale as the measurements, which affected the stability and reproducability of the photocurrent, and made repositioning the transmitter and receiver challenging. In Version 2, we have added high precision locking stages with custom clamps to better hold the alignment constant over time.

\paragraph{Versatility for different setups} FLS Version 1 was designed to be mounted on a stage above a receiver, such as the LAT optics tube test receiver, with coupling optics directing the laser beam downward into the receiver. As a result, the beam could only be directed downward. However, flexibility in the direction of the FLS output is critical for coupling to different test setups. To accommodate a wider variety of coupling setups, FLS Version 2 is designed with openings at the top, bottom, and side of the instrument, with options to mount coupling optics to any of those openings.

\paragraph{External casing} For use in the field, we require protection from environmental factors like dust. Dust that is able to reach the ends of the fiber optics can affect our measurement capabilities, so some shielding of the TeraScan 1550 elements that use fiber optic cables is particularly important. Additionally, shielding the telescope from stray light from the FLS setup itself is also critical for in-field measurements. Improving on the design from Version 1, which did not have a full enclosure, we designed a case for Version 2 using T-slot framing to enclose the internal module and sensitive electronics of the FLS.

\subsection{Procedural Improvements}\label{subsec:procedures}
\paragraph{Laser warm-up} The absolute frequency calibration is made by calibrating off of atmospheric water vapor absorption lines within the frequency range of the laser system. Using the Version 1 prototype, we observed a drift in the TeraScan 1550 laser frequency over time following a negative exponential curve\cite{shreya-fls-optics-2026}. Based on those observations, we concluded that the laser system requires at least 24 hours to warm up so that it can reach a stable frequency. Once the system is warmed up, the frequency is stable and repeatable, as is discussed in Section \ref{subsec:frequency-drift}. We measured this warm-up time on two independent TeraScan 1550 systems, indicating that all future FLS measurements using a TeraScan 1550 system must allow for this warm-up time.

\paragraph{Data acquisition} The TeraScan 1550 system is ordinarily operated using Toptica's proprietary TOPAS TeraScan Control Software\cite{toptica_terascan1550}, which is a Graphical User Interface (GUI) program intended to allow the user to connect to the laser system, change the effective frequency of the laser, and measure the photocurrent in the system's receiver. However, the TOPAS software does not allow the user to record the timestamps for measurements, which is a requirement for synchronizing measurements of the laser frequency with detector responses. In measurements with FLS Version 1, we used a Python interface to send commands, read values, and record timestamps from the FLS control computer; however, the interface did not allow synchronizing timestamps across computers and did not offer the full functionality of the GUI. For Version 2, we developed more robust data acquisition and control software that can be synchronized with the detectors by developing an Agent within the Observatory Control Software (OCS) framework\cite{koopman2020ocs}.

\section{System Characterization}\label{sec:characterization}

Tests with FLS Version 2 use a different Toptica TeraScan 1550 system than those with Version 1  due to the availability of different systems at the time of testing. The system used in Version 2 does not have a phase modulation extension, reducing its spectral resolution, and it has a range up to 880 GHz, which allows for calibrating off of the 557 GHz and 775 GHz water vapor absorption lines. Improved source characterization can be achieved by subsituting a different TeraScan 1550 system into the designed Version 2 FLS with the phase modulation extension and a higher upper frequency limit to enable measuring additional atmospheric lines, as was done for Version 1 tests described in Sutariya (2026)\cite{shreya-fls-optics-2026}.

\subsection{Photocurrent reference measurement}\label{subsec:photocurrent-reference}
\begin{figure}
    \centering
    \includegraphics[width=0.8\linewidth]{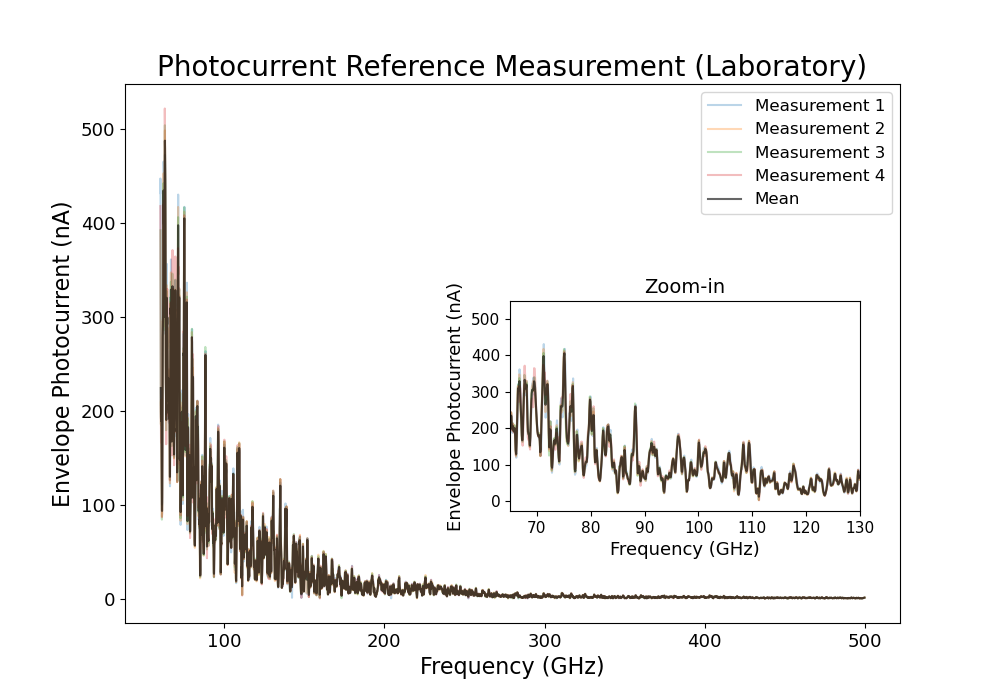}
    \caption{Measurements of the photocurrent using the TeraScan 1550 receiver photomixer in the laboratory show a repeatable pattern of oscillatory behavior. Because the response is consistent, we can correct for this behavior.}

    \label{fig:lab-photocurrent}
\end{figure}

The Toptica TeraScan 1550 transmitter has a varying power output with frequency that is consistent and repeatable between measurements when the alignment is held constant. This power output can be observed as the photocurrent measured in the receiver photodiode. To ensure that we account for this variation when measuring detector responses with the Version 2 FLS, we measured the photocurrent with the receiver photodiode in the laboratory setting with only mirrors installed both within and outside of expected instrument passbands for CMB experiments. We made multiple measurements of this power variation in the self-calibration configuration over the course of a week to confirm the consistency of our measurements and to ensure that this calibration was repeatable. These measurements are shown in Figure \ref{fig:lab-photocurrent}.

\subsection{Prism reflectivity}\label{subsec:prism-reflectivity}
\begin{figure}
    \centering
    \includegraphics[width=\linewidth]{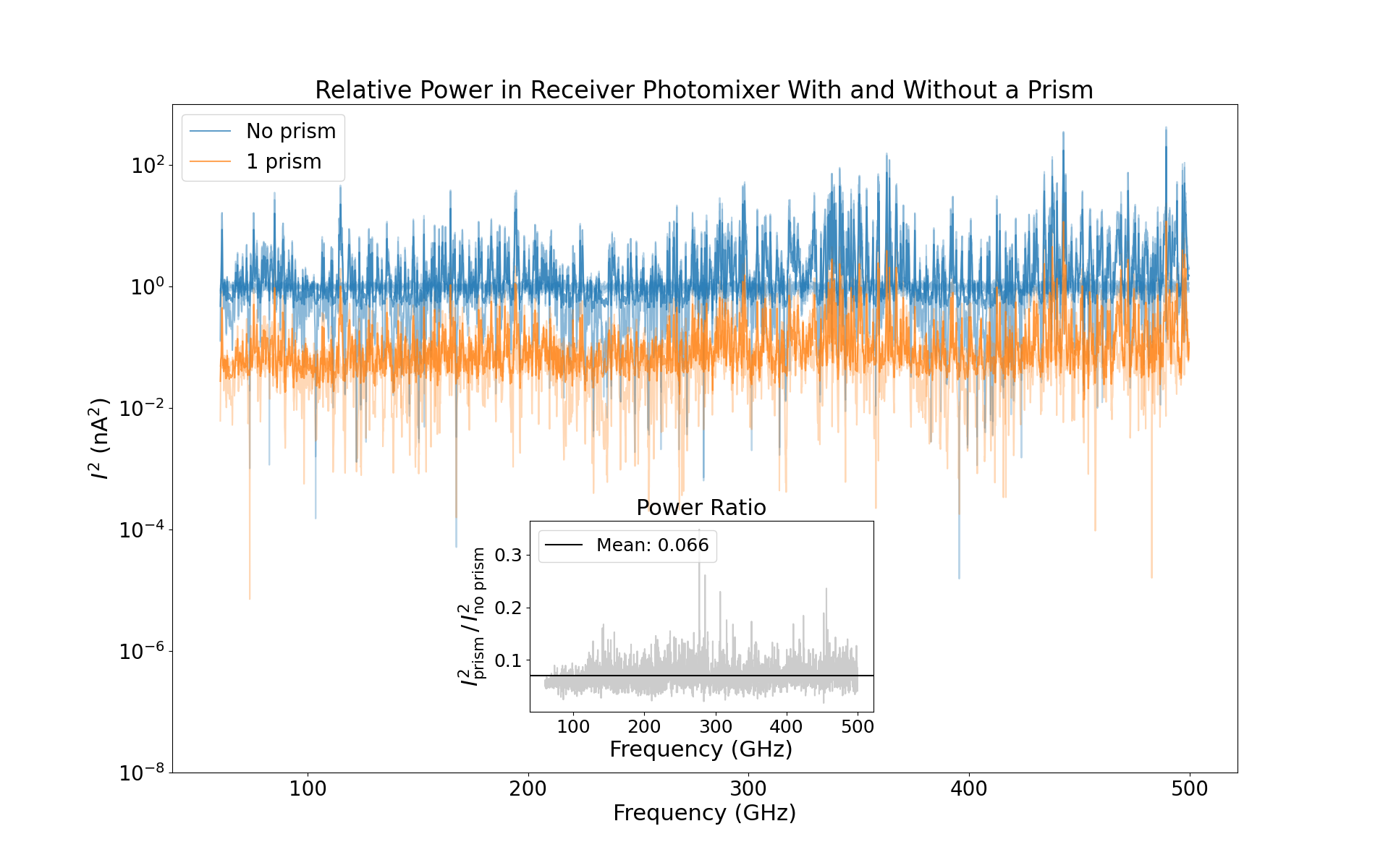}
    \caption{Measurements in the self-calibration setup were made with and without a prism installed to characterize the reflectivity of the prisms. The individual measurements without a prism are shown in lighter blue, with the average of the measurements shown in darker blue. Likewise, the individual measurements with one prism installed are shown in lighter orange, with the average of the measurements shown in darker orange. After correcting for the photocurrent reference, we compared the ratio of the power incident on the receiver photomixer with one prism to the power with no prisms installed, finding that the prisms reflects 6.6\% of the power (93.4\% attenuation), and that the attenuation is neutral across all measured frequencies.}
    \label{fig:reflectivity}
\end{figure}

For the Version 2 FLS, we fabricated new Nylon 6/6 attenuating prisms to ensure that the size and hole pattern of the prisms matched the new design. We then characterized the attenuation of the prisms to ensure that it was neutral across frequencies and to account for any variation in optical properties between material batches. To do so, we first took baseline measurements with only mirror flats installed, where the transmitter and receiver photomixers were installed on the micrometer stages and aligned before running a frequency scan between 65 GHz and 500 GHz, with both increasing and decreasing frequency to account for any hysteresis in the measurements. We then removed the first mirror panel and installed a prism in its place, and repeated the same frequency scans. This procedure was repeated several hours later. We compared the power incident on the receiver photomixer with and without the prism installed, as shown in Figure \ref{fig:reflectivity}, and found that each prism attenuates 93.4\% of the signal, making the new prisms comparable to the Version 1 FLS prisms. The attenuation is also measured to be neutral across the frequency range.

\subsection{Frequency drift over time}\label{subsec:frequency-drift}
\begin{figure}
    \centering
    \includegraphics[width=0.8\linewidth]{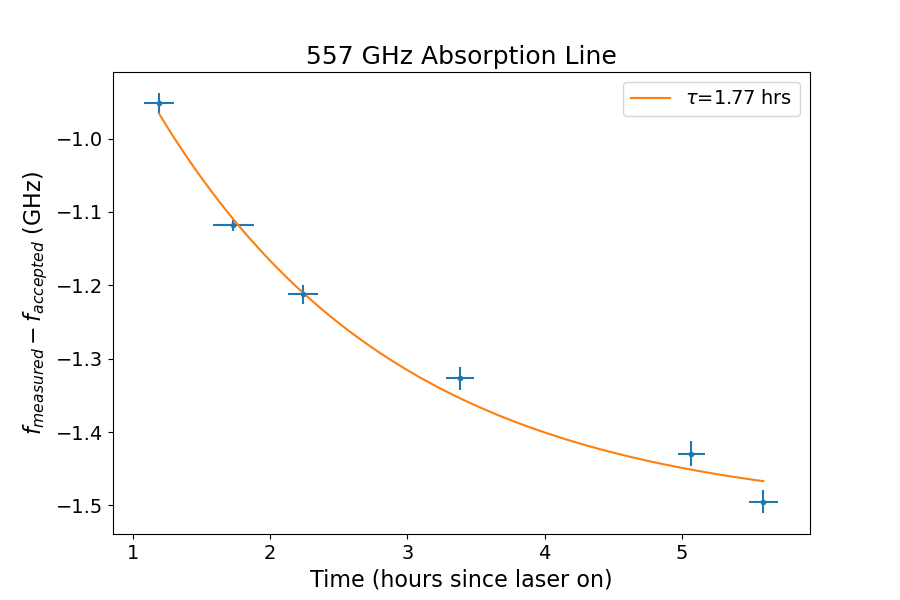}
    \caption{An example of measurements of the center of the 557 GHz atmospheric water vapor absorption line shows that the frequency of the laser changes over time, with a time constant of approximately 1.77 hours. This measurement was repeated with both the 557 GHz and 775 GHz water vapor absorption lines to check the consistency of the TeraScan system's behavior. This is a similar pattern to the one seen when characterizing the Toptica TeraScan 1550 system in Sutariya (2026)\cite{shreya-fls-optics-2026}, and shows that we need to wait for the laser to warm up before we begin taking measurements.}
    \label{fig:frequency-drift-line1}
\end{figure}

In our prior work with the Version 1 prototype, we determined that there is a drift in the frequency of the laser over time after turning the Toptica TeraScan 1550 system on, which follows an exponential decay pattern\cite{shreya-fls-optics-2026}. Due to this observation with Version 1, we followed the same water vapor line analysis with the laser used in the Version 2 prototype to determine a safe settling time before taking measurements with this laser. For these water vapor line measurements, we installed the Toptica TeraScan 1550 system on the FLS Version 2 structure with only mirrors installed, and used the locking micrometer stages for optical alignment. After allowing the laser system to warm up for 1 hour to reach full power, we began taking measurements of the water vapor absorption lines at 557 GHz and 775 GHz, alternating between sweeps with increasing and decreasing frequency. For each measurement time, we took 5 measurements of each water vapor absorption line. We then made Gaussian fits of the water vapor absorption lines to determine the observed center frequencies. This procedure was repeated after turning the laser off for a few days. An example measurement of the 557 GHz water vapor absorption line is shown in Figure \ref{fig:frequency-drift-line1}, which shows the exponential decay that we observed with the Version 2 laser. Due to measurement uncertainties, the expected time constants based on measurements of both water vapor absorption lines ranged from 1.75 hours to 2.53 hours; to ensure that the frequency is stable for detector measurements, we choose to base our procedures on the longer time constant, allowing a minimum of 24 hours for the laser frequency to stabilize before taking measurements. We have demonstrated that after this settling time, the laser frequency is stable and frequency measurements between power cycles are consistent.

\subsection{Frequency calibration}\label{subsec:frequency-calibration}
\begin{figure}
    \centering
    \includegraphics[width=0.7\linewidth]{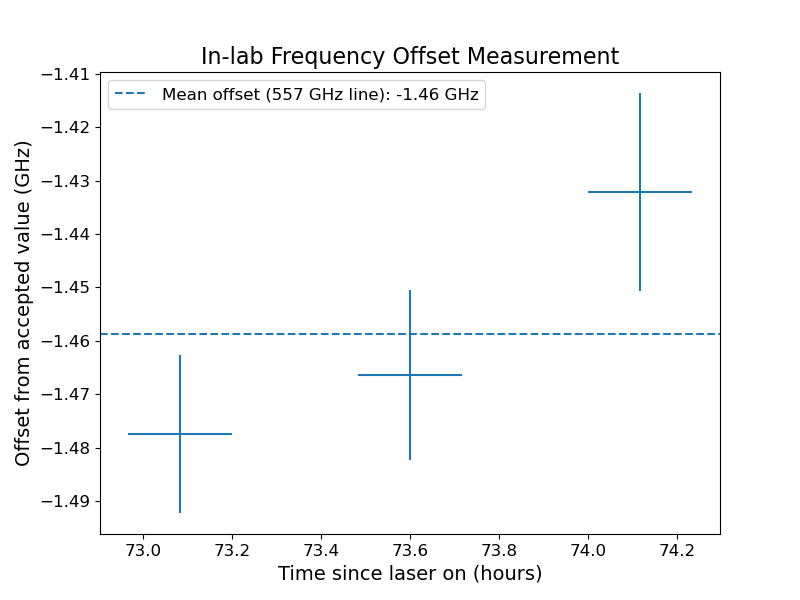}
    \caption{An example of the frequency offset measurement performed in the laboratory to characterize the TeraScan 1550 laser system. After allowing the laser system to remain on for over 72 hours, we scanned across frequency ranges that encompassed the expected water vapor absorption lines. We then analyzed the data to find the centers of the water vapor absorption lines as measured by the FLS. The analysis shows that the measured central frequencies of the water vapor absorption lines are lower than the expected values; repeated measurements place this offset at $-1.5 \pm 0.1$ GHz, which we treat as an additive constant.}
    \label{fig:freq_offset}
\end{figure}

Measurements of the atmospheric water vapor absorption lines with the TeraScan 1550 system used in FLS Version 1 showed that the true frequency was offset from the intended frequency\cite{shreya-fls-optics-2026}, so we calibrated the system used in Version 2 tests to characterize any offset in the same way. After installing the laser system in the FLS structure with only mirrors installed, we powered the system on and allowed it to remain on for 72 hours to ensure that the frequency was stable. We then scanned across frequency ranges that included the water vapor absorption lines at 558 GHz and 755 GHz, and analyzed those data to find the measured central frequencies of the absorption lines. This procedure was repeated several times on the same day, and was also repeated on a later day after power cycling the lasers and allowing them to settle again. An example measurement is shown in Figure \ref{fig:freq_offset}. Combining all data, we found that the frequency of the laser was offset by $-1.5 \pm 0.1$ GHz, which we treat as a constant offset. This value is used to correct the frequencies in any measurements we make with the FLS.

\section{Field Testing with the Simons Observatory Large Aperture Telescope}\label{sec:so-lat}
We performed the first ever field test of this new technology with FLS Version 2 by deploying it to the Simons Observatory (SO) in the Atacama Desert in Chile. Our tests involved coupling the FLS to an optics tube on the SO Large Aperture Telescope (LAT)\cite{LAT_2021} and measuring the frequency response of the mid-frequency (MF) bands containing 90 GHz and 150 GHz detectors\cite{mf_ufps}. This section details the optical coupling designed for use with the SO LAT and the integration of the FLS software into OCS for use at the SO site. We additionally include measurements of the LAT MF spectral response.

\subsection{Optical coupling}\label{subsec:coupling}
\begin{figure}
    \centering
    \includegraphics[ width=0.8\linewidth]{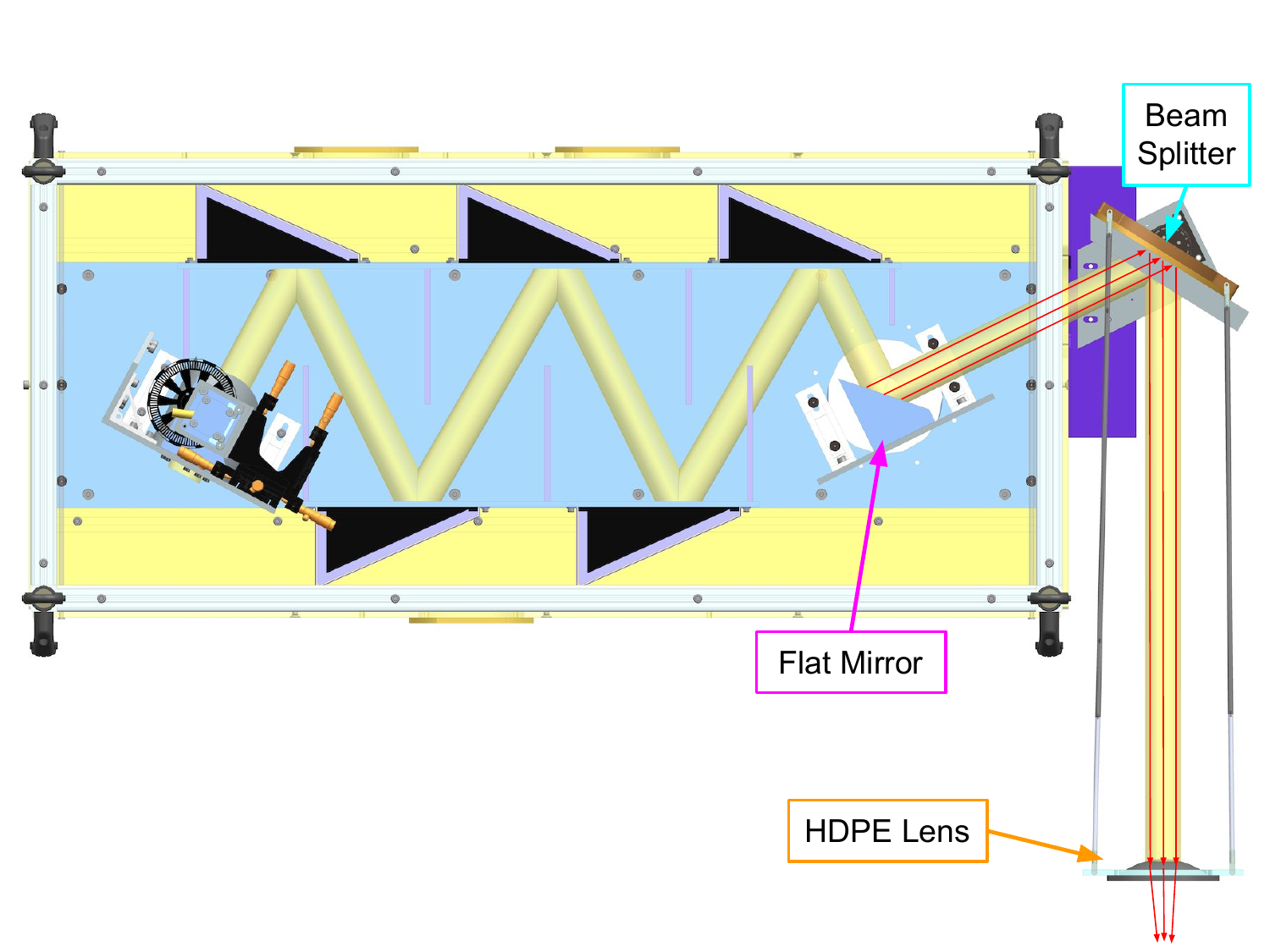}
    \caption{A drawing of the optical coupling apparatus used to couple the FLS to the SO LAT optics tube installed on the FLS. The laser is reflected out of the FLS by a $45^\circ$ flat mirror. It is then reflected off of a beam splitter, and goes through a lens that couples it to the detectors in the LAT optics tube. Arrows trace the path of the laser after reflecting off of the $45^\circ$ flat mirror, showing that the beam remains collimated until it reaches the lens, which focuses the beam into the receiver. An NDF (not pictured) is placed directly in front of the LAT optics tube window.}
    \label{fig:so-lat-coupling}
\end{figure}

To ensure that the FLS is properly coupled to detectors in the SO LAT receiver, we required coupling optics that would focus the laser on the detectors. For the in-field tests, we used the side port of the FLS to output the signal. The OAP mirror on the output side was replaced by a flat mirror to output a collimated beam. The beam is directed through an angled beam splitter made of tensioned polypropylene to reduce the optical loading and to direct the output to the LAT receiver. The beam then passes through a converging lens made of HDPE to focus the FLS beam onto the detectors. A neutral density filter (NDF) is used to further attenuate power at the telescope window. A drawing of this optical coupling system is shown in Figure \ref{fig:so-lat-coupling}.

\subsection{Software integration into OCS}\label{subsec:ocs-agent}
As mentioned in Section \ref{subsec:procedures}, developing improved data acquisition and control software for the FLS is critical for use of the FLS with detectors. In alignment with SO's fielded software system, we developed an OCS Agent\cite{koopman2020ocs} to support the FLS. OCS uses a central timing system to synchronize the timestamps of data collected across different Agents and computing nodes, allowing us to synchronize the FLS timestamps with those of the detector timestreams. The full functionality required for the FLS measurement procedures was available via a TCP command line interface. Multiple experiments, including SO\cite{koopman-deployment} and SPT-SLIM\cite{slim-performance} use implementations of OCS, so the Agent developed for field testing with SO is broadly usable across both in-lab and in-field platforms.

The Agent allows the following functionality:
\begin{itemize}
    \item Open a connection to the Toptica TeraScan 1550 system via a TCP interface.
    \item Power the lasers of the TeraScan 1550 system on and off.
    \item Set the TeraScan 1550 bias amplitude voltage and bias offset voltage to be either zero or the default values set by the manufacturer.
    \item Set the effective frequency of the laser system.
    \item Command the TeraScan 1550 system to perform a ``fast'' frequency sweep between a minimum and maximum frequency value. The user may also control the step size between frequencies, the integration time at each frequency step, and the direction of the sweep (increasing frequency or decreasing frequency).
    \item Command the TeraScan 1550 system to stop a frequency sweep.
    \item Perform continuous data acquisition, recording the set frequency, actual frequency measured by the TeraScan 1550 system, photocurrent measured by the receiver photomixer, laser state (on/off), bias amplitude voltage, bias offset voltage, sweep parameters, and timestamps. The data are both saved to file in the OCS \texttt{.g3} file format and passed to InfluxDB, which is used as a backend for live monitoring via Grafana, a web-based graphical interface\cite{koopman2020ocs}.
\end{itemize}

\subsection{Measurements of the LAT spectra}\label{subsec:lat-measurements}
\begin{figure}
    \centering
    \includegraphics[width=\linewidth]{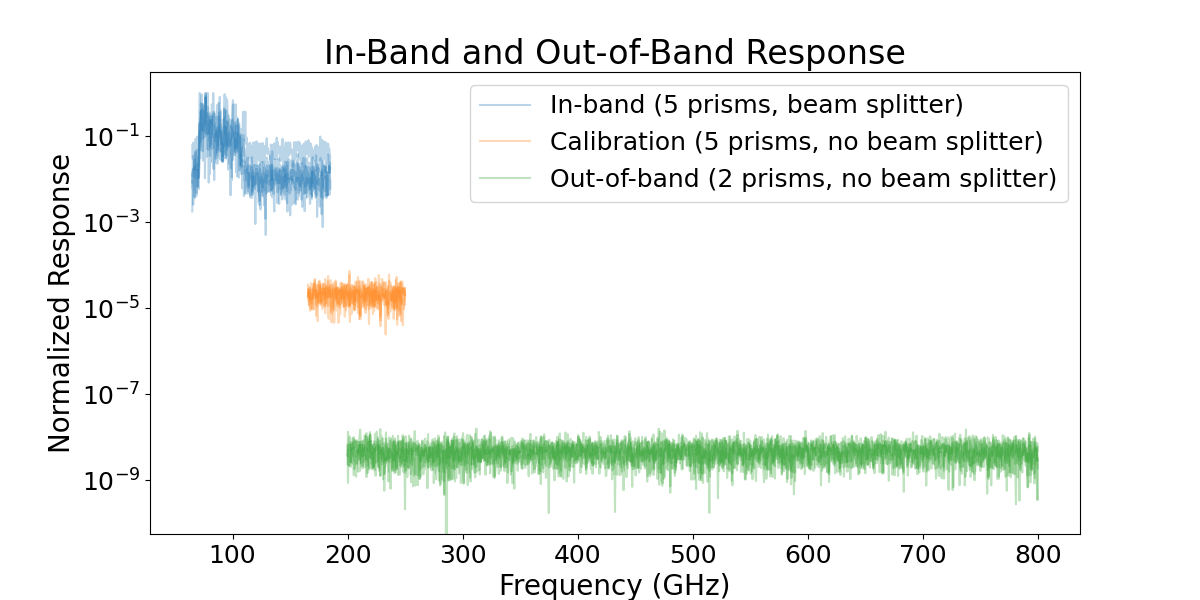}
    \caption{A sample of five detectors show a 90 GHz band response to the FLS laser with 5 prisms installed (blue) compared with the out-of-band response with 2 prisms installed in the FLS and a mirror in place of the beam splitter (green), normalized to the peak of the in-band response. We also include a calibration scan with 5 prisms installed in the FLS and a mirror in place of the beam splitter (orange). We do not see a significant response outside of the band, indicating that there is no high-frequency response leakage up to 800 GHz.}
    \label{fig:outofband}
\end{figure}

Initial measurements of the SO LAT detector spectra were taken May 15-19, 2026. After initially setting up the TeraScan 1550 system, the lasers were allowed to warm up for 36 hours to ensure frequency stability during measurements with the LAT detectors. The FLS housing was then assembled with the TeraScan 1550 system installed in it, and the coupling optics were installed and aligned with the LAT optics tube window. Tests with the chopped FLS signal were used to determine the number of prisms required for in-band measurements, and an on/off signal test performed by covering the FLS output port was performed to ensure that the chopped signal measured was the FLS signal. In-band measurements consisted of scanning between 65 GHz and 185 GHz after biasing the detectors; some measurements were taken with 4 prisms and one flat mirror installed in the FLS, while others were taken with 5 prisms to further reduce the power from the laser and avoid saturating the detectors. The raw detector signals were then demodulated using the chopper frequency. Because not all of the detectors on a module were illuminated by the FLS, only detectors with a signal peak at least 5 times the noise floor were included in our analysis. On order $\sim$1000 detectors responded to the FLS signal; a subset of those detector responses in the 90 GHz band with 5 prisms is shown in Figure \ref{fig:outofband}. The NDF used has been measured to have a constant response across all the frequencies we measured, so no correction is applied. The measured band edges are consistent with previous in-lab FTS measurements at the 1 GHz level. Future analysis of the bandpasses will include accounting for optical effects from the lens and beam splitter and accounting for the calibrated output of the transmitter.

In addition to measuring the in-band response, the FLS is well-suited to search for out-of-band detector responses, as the power incident on the detectors may be increased by reducing the number of attenuating prisms installed in the FLS and replacing them with mirrors. To perform the out-of-band measurements of the SO LAT detectors, we scanned between 200 GHz and 800 GHz. We ran versions of this measurement with 5 prisms and the beam splitter to match the in-band measurements, and with 2 prisms and a flat mirror in place of the beam splitter to increase the incident power by $\sim$$1\times10^5$. The measurements of a subset of detectors for the 2 prism and mirror configuration are shown in Figure \ref{fig:outofband}. The out-of-band signal is consistent with noise and shows no structure. We are thus able to constrain the out-of-band leakage to be below on order $10^{-5}$ in both the 90 GHz and 150 GHz bands, which is the most stringent constraint to date.

\section{Discussion and future measurement plans}\label{sec:discussion}
The first successful in-field measurement with an FLS of the SO LAT detector spectra in the 90 GHz and 150 GHz bands demonstrates the viability of using this novel calibrator for bandpass measurements both in the lab and in the field. The out-of-band constraints are the strongest constraints placed on out-of-band leakage on the SO LAT to date, demonstrating the power of this new calibration method for constraining out-of-band leakage. Future measurements plan to use fully reflective coupling optics to reduce the optical loading, eliminating the need for an NDF, and enable measuring more detectors at once. Reflective optics for the SO FTS system are currently in production and could also be used for future FLS measurements. Additionally, the FLS is expected to provide further improved constraints on detector bandpass when used in combination with the FTS. In future measurements, we plan to compare FLS and FTS measurements to assess their performance differences and better characterize the performance of both systems.

%%%%%%%%%%%%%%%%%%%%%%%%%%%%%%%%%%%%%%%%%%%%%%%%%%%%%%

\acknowledgements
\begin{sloppypar}
This manuscript has been authored by Fermi Forward Discovery Group, LLC under Contract No. 89243024CSC000002 with the U.S. Department of Energy, Office of Science, Office of High Energy Physics. This work was supported through a Fermilab Laboratory Directed Research and Development (LDRD) award. This work was supported in part by a grant from the Simons Foundation (Award \#457687, B.K.). This work was supported by the U.S. National Science Foundation (Award Number: 2153201).
\end{sloppypar}

\bibliographystyle{spiebib}
\bibliography{bibliography.bib}

\end{document}